\documentclass[journal=apchd5,manuscript=article]{achemso}

\usepackage{chemformula} 
\usepackage[T1]{fontenc} 
\usepackage{physics}
\usepackage{subcaption}
\usepackage{hyperref}
\usepackage{amssymb}

\makeatletter
\renewcommand*{\acs@contact@details}{%
  {\sffamily Corresponding authors: \acs@email@list}%
  \acs@number@list
}
\makeatother

\author{Stavros Athanasiou}
\affiliation[EPFL]
{Nanophotonics and Metrology Laboratory (NAM), Swiss Federal Institute of Technology Lausanne (EPFL), 1015 Lausanne, Switzerland}
\alsoaffiliation
{Departamento de Física Teórica de la Materia Condensada, Universidad Autónoma de Madrid, E-28049 Madrid, Spain}
\alsoaffiliation
{Condensed Matter Physics Center (IFIMAC), Universidad Autónoma de Madrid, E-28049 Madrid, Spain}
\email{stavros.athanasiou@epfl.ch}

\author{Olivier J.F.\ Martin}
\affiliation[EPFL]
{Nanophotonics and Metrology Laboratory (NAM), Swiss Federal Institute of Technology Lausanne (EPFL), 1015 Lausanne, Switzerland}

\author{Francisco J. García-Vidal}
\affiliation[UAM]
{Departamento de Física Teórica de la Materia Condensada, Universidad Autónoma de Madrid, E-28049 Madrid, Spain}
\alsoaffiliation
{Condensed Matter Physics Center (IFIMAC), Universidad Autónoma de Madrid, E-28049 Madrid, Spain}

\author{Johannes Feist}
\affiliation[UAM]
{Departamento de Física Teórica de la Materia Condensada, Universidad Autónoma de Madrid, E-28049 Madrid, Spain}
\alsoaffiliation
{Condensed Matter Physics Center (IFIMAC), Universidad Autónoma de Madrid, E-28049 Madrid, Spain}
\email{johannes.feist@uam.es}

\title[]{
    Hyperbolic Nanocavities for Single-Molecule Vibrational Strong Coupling
}

\usepackage{xcolor}

\usepackage{todonotes}

\definecolor{linkcolor}{cmyk}{0,0.87,0.68,0.32}
\definecolor{citecolor}{cmyk}{0.94,0.54,0,0}
\definecolor{urlcolor}{cmyk}{1,0.50,0,0}
\hypersetup{colorlinks=true, linkcolor=linkcolor, citecolor=citecolor, urlcolor=urlcolor}

\SectionNumbersOn

\begin{document}



\begin{abstract}
    Hyperbolic phonon polaritons (HPhPs) in low-loss anisotropic materials such as hexagonal boron nitride (hBN) provide extreme sub-diffractional field confinement in the mid-infrared, making them attractive for enhancing the interaction of light with molecular vibrations. Collective vibrational strong coupling between molecular ensembles and HPhPs has already been demonstrated in extended slabs. Reaching this regime with a single molecule, however, remains a major challenge, and would provide a platform for investigating how vacuum electromagnetic fields influence molecular dynamics and chemical reactivity. Here, we show theoretically that single-molecule vibrational strong coupling can be reached in hBN nanocavities that laterally confine HPhPs within a finite volume. Combining full-wave electromagnetic simulations with the few-mode quantization approach, we find that these nanocavities support a rich set of asymmetric resonances that strongly enhance the local spectral density, with frequencies that follow from simple HPhP ray trajectories and can be tuned through the cavity aspect ratio. We identify resonances with quality factors of about $600$ and effective mode volumes of $|V_{\mathrm{eff}}| \sim 10^{-7}\lambda^{3}$, sufficient for single-molecule strong coupling with vibrational transitions with dipole moments on the order of $1\,\mathrm{D}$. These results establish hyperbolic nanocavities as a promising platform for single-molecule vibrational strong coupling and open new routes toward the local control of vibrational dynamics and molecular energy flow.
\end{abstract}




\section{Introduction}
\label{sec:intro}

Strong light-matter coupling is reached when the coherent exchange of energy between an optical mode and a material excitation becomes faster than the relevant dissipative processes, giving rise to hybrid polaritonic states with properties inherited from both constituents~\cite{Garcia-Vidal2021}. It has been demonstrated across a broad range of platforms, including atoms and quantum dots in optical cavities~\cite{Aoki2006,Hennessy2007}, superconducting circuits~\cite{Wallraff2004}, and molecules~\cite{Chikkaraddy2016,Menghrajani2024}. Beyond the characteristic normal-mode splitting, strong coupling can qualitatively modify excitation transport~\cite{Feist2015,Schachenmayer2015} and give rise to complex many-body dynamics~\cite{FowlerWright2022}. For molecules, the coupling of an individual emitter to a cavity mode is typically weak, and strong coupling is therefore usually reached collectively, with $N$ molecules coupled to the same mode enhancing the coupling strength by a factor $\sqrt{N}$. Strong coupling of a single molecule is substantially more demanding, and has been achieved for electronic transitions in plasmonic nanocavities with extreme field confinement~\cite{Chikkaraddy2016}.

Molecular vibrations in the mid-infrared can likewise be strongly coupled to optical modes, leading to vibrational strong coupling (VSC). Collective VSC has been demonstrated in Fabry-Perot cavities~\cite{Long2015,Shalabney2015}, with surface phonon polaritons in quartz micropillars~\cite{Liu2022}, and in plasmonic cavities~\cite{Lather2025}, and has been found to modify intramolecular vibrational energy redistribution~\cite{Chen2022Cavity} and energy transfer~\cite{Xiang2020,Brawley2025}. Changes of ground-state chemical reactivity under collective VSC have also been reported~\cite{Thomas2016,Thomas2019,Ahn2023}, but their microscopic origin remains under active investigation~\cite{Galego2019,Xiang2024}. A central open question is how collective coupling, which is distributed over a macroscopic number of molecules, translates into modifications of the individual molecules. Reaching VSC at the single-molecule level would make it possible to study such local effects directly and to distinguish them from collective phenomena. It would furthermore enable quantum optics with individual molecular vibrations in the mid-infrared, a spectral range in which quantum-optical sources, detectors, and integrated components remain comparatively underdeveloped.

Phonon polaritons provide a natural platform for reaching this regime. They arise from the coupling of electromagnetic fields to optical phonons in polar crystals. In anisotropic materials such as hexagonal boron nitride (hBN), the permittivity components along different crystal axes can have opposite signs, giving rise to hyperbolic phonon polaritons (HPhPs)~\cite{Dai2014,Jacob2014}. These modes support wave vectors far exceeding the free-space one, enabling deeply sub-diffractional confinement, and propagate as highly directional rays~\cite{Dai2014,Jacob2014,Ma2018,Wang2024}. Together with their comparatively low losses~\cite{Ma2018,Giles2018}, this has enabled a broad range of nanophotonic applications, including sub-diffractional imaging and focusing~\cite{Li2015}, cooling of graphene devices by a hyperbolic substrate~\cite{Yang2018}, electroluminescence mediated by HPhPs~\cite{Guo2025,Abou-Hamdan2025}, and fast mid-infrared photodetection with plasmonic antennas and graphene~\cite{Castilla2020}.

Collective VSC between organic molecules and HPhPs has indeed been observed in extended hBN and $\alpha\text{-MoO}_3$ slabs~\cite{Bylinkin2021,Tresguerres-Mata2025}. Extended slabs alone, however, are not sufficient to reach the single-molecule regime. While they confine the field strongly in the out-of-plane direction, HPhPs propagate freely in the plane, so that a molecule couples to a broad continuum of propagating modes. This strongly enhances its decay rate but does not produce the discrete, spectrally isolated resonances required for coherent energy exchange with a single molecular vibration. Lateral confinement in a finite resonator provides a natural route to overcome this limitation: the propagating HPhPs are reflected back toward the emitter instead of carrying energy away from it, and their constructive interference forms discrete resonances.

Finite hyperbolic structures supporting localized polaritonic resonances have been realized in a variety of settings, including hBN metasurfaces~\cite{Li2018}, high-quality resonators based on $\alpha$-MoO$_3$ nanoribbons~\cite{Yu2022}, and hyperbolic metamaterial nanocavities with ultrasmall mode volumes in the visible~\cite{Indukuri2019}. Lateral HPhP confinement has also been achieved without patterning the hyperbolic material itself, by placing an extended hBN slab above an aperture in a gold substrate, where the contrast in the local polaritonic dispersion confines the HPhPs within the aperture region~\cite{HerzigSheinfux2024}. However, the potential of finite hyperbolic resonators for reaching VSC at the single-molecule level remains largely unexplored.

In this work, we show theoretically that hBN nanocavities, which laterally confine HPhP propagation within a finite volume, enable VSC with a single molecule. We compute the spectral density experienced by a molecular vibration close to the nanocavity using full-wave electromagnetic simulations, and analyze the hybrid system using the few-mode quantization approach~\cite{Medina2021,Monica2022}. This approach encodes the dispersive and lossy electromagnetic environment described by macroscopic QED~\cite{Scheel2009} into a small set of lossy modes, providing a master-equation description of the dynamics and direct access to the figures of merit governing the interaction. We find that the nanocavities support pronounced resonances with high quality factors and extremely subwavelength mode volumes, whose frequencies are set by simple HPhP ray trajectories and can be tuned through the cavity aspect ratio.

The article is organized as follows. \autoref{sec:theory} introduces the theoretical and numerical framework, and \autoref{sec:planar} reviews the HPhP ray picture and the emitter spectral density in planar hBN\@. In \autoref{sec:cavities}, we introduce the hyperbolic nanocavities, analyze their resonances using the few-mode model, and demonstrate single-molecule VSC. \autoref{sec:size} examines how the interaction can be controlled through the cavity dimensions, including an ultrasmall cavity in which the simple ray picture breaks down. We discuss experimental requirements and the role of the few-mode description in \autoref{sec:discussion}, and conclude in \autoref{sec:conclusions}.



\section{Theoretical and Numerical Framework}
\label{sec:theory}

The interaction of a quantum emitter at position $\mathbf{r}$ and transition dipole moment $\boldsymbol{\mu} = \mu \hat{\mathbf{n}}$ with its electromagnetic environment is fully characterized by the spectral density~\cite{Novotny2012}
\begin{equation}
    J(\mathbf{r},\omega) = \frac{\mu^2\omega^2}{\pi \hbar \varepsilon_0 c^2} \hat{\mathbf{n}} \cdot \Im\left[ \mathbf{G}(\mathbf{r},\mathbf{r},\omega) \right] \cdot \hat{\mathbf{n}},
\end{equation}
where $\mathbf{G}(\mathbf{r},\mathbf{r},\omega)$ is the dyadic electromagnetic Green's function evaluated at coincident points. For planar geometries, $\mathbf{G}$ can be obtained semi-analytically from its plane-wave (Weyl) expansion in the $xy$-plane (Supplementary Note~S1). For the nanocavities, we compute the spectral density numerically with the finite-element package COMSOL Multiphysics (Supplementary Note~S3).

Within macroscopic quantum electrodynamics, the electromagnetic field in a dispersive and absorbing environment is described by a four-dimensional continuum of bosonic excitations, labeled by position and frequency (and a discrete polarization index), whose collective effect on the emitter dynamics is fully determined by $J(\mathbf{r},\omega)$~\cite{Scheel2009}. While formally exact, this description contains an infinite number of electromagnetic degrees of freedom. The few-mode quantization framework~\cite{Medina2021,Monica2022} maps it onto an effective model consisting of a finite number of interacting bosonic modes, each coupled to a Markovian reservoir, which are chosen such that they reproduce the spectral density within a selected frequency window~\cite{Dalton2001,Pleasance2020}. The reduced density matrix $\hat\rho$ of the emitter and the effective modes then obeys the Lindblad master equation (here and in the following, $\hbar = 1$)
\begin{equation}
    \label{eq:meqtime}
    \dot{\hat{\rho}} = - i [\hat{H}_\text{S},\hat{\rho}] +
    L_{e}[\hat{\rho}] +
    \sum_i L_{a_i}[\hat{\rho}],
\end{equation}
where $L_{k}[\hat{\rho}] = \hat{\mathcal{O}}_k \hat{\rho} \hat{\mathcal{O}}_k^\dagger - \frac{1}{2} \{ \hat{\mathcal{O}}_k^\dagger \hat{\mathcal{O}}_k,\hat{\rho} \}$ is a Lindblad dissipator with jump operators $\hat{\mathcal{O}}_e = \sqrt{\gamma_e} \hat{b}$ for the emitter and $\hat{\mathcal{O}}_{a_i} = \sqrt{\kappa_i} \hat{a}_i$ for the effective modes. The system Hamiltonian is
\begin{equation}
    \hat{H}_\text{S} = \omega_e \hat{b}^\dagger \hat{b} + \sum_{i,j} \omega_{ij} \hat{a}^\dagger_{i} \hat{a}_j + \mu \sum_i g_i (\hat{b}^\dagger + \hat{b}) ( \hat{a}_i + \hat{a}^\dagger_i).
\end{equation}
Here, the molecular vibration is described as a harmonic oscillator with annihilation operator $\hat b$, frequency $\omega_e$, and intrinsic non-radiative decay rate $\gamma_e$. The operators $\hat a_i$ annihilate excitations in the effective modes, whose energies and mutual couplings are given by $\omega_{ij}$ and whose decay rates are $\kappa_i$. The couplings $g_i$ to the emitter have units of electric field. The spectral density of this model is~\cite{Medina2021}
\begin{equation}
    \label{eq:Jfit}
    J_\text{fm}(\omega) = \frac{\mu^2}{\pi} \text{Im}\left\{ \vec{g}^T\frac{1}{\mathbf{H}-\omega} \vec{g} \right\},
\end{equation}
where $\mathbf H_{ij}=\omega_{ij}-\frac{i}{2}\kappa_i\delta_{ij}$ is the non-Hermitian mode matrix. The model parameters are obtained by fitting \autoref{eq:Jfit} to the spectral density computed from the classical Green's function.

Since we are interested in strong coupling to a specific resonance, the fit is performed only within a frequency window around the emitter transition frequency, which contains the modes responsible for non-Markovian dynamics. The remaining part of the spectral density is treated perturbatively~\cite{Carlos2024}: it produces a Lamb shift, described by $\hat H_{\mathrm{pert}}=\Delta_{\mathrm{pert}}\hat{b}^\dagger \hat{b}$, and an additional emitter decay channel with jump operator $\hat{\mathcal{O}}_{e, \text{pert}} = \sqrt{\gamma_{\mathrm{pert}}} \hat{b}$ (Supplementary Note~S4.2).

Molecular vibrations can be noticeably anharmonic~\cite{Kalarde2025}. Here, we restrict ourselves to linear response and treat the light-matter interaction within the rotating-wave approximation (RWA), which is well justified since all coupling strengths are several orders of magnitude smaller than $\omega_e$. The RWA conserves the total excitation number, such that the dynamics starting from a single excitation remain within the zero- and single-excitation sectors. Within these sectors, the harmonic oscillator is equivalent to an anharmonic vibration or a two-level system, independently of the magnitude of the anharmonicity (Supplementary Note~S4.3). Finally, we neglect thermal excitation of the effective modes and treat their reservoirs as zero-temperature baths, since at the relevant energies, $\hbar\omega\approx0.17$--$0.2\,\mathrm{eV}$, the thermal occupation is on the order of $10^{-3}$ or below even at room temperature.

Diagonalizing the non-Hermitian mode matrix $\mathbf{H}$ by a complex orthogonal transformation (Supplementary Note~S4.1) yields effective uncoupled modes with complex frequencies $\tilde\Omega_i$ and transformed couplings $\tilde g_i$. Their quality factors are
\begin{equation}
    Q_i = - \frac{ \mathrm{Re}\{ \tilde\Omega_i \} }{2 \mathrm{Im} \{ \tilde\Omega_i \} },
\end{equation}
and the physical light-matter coupling strength of mode $i$ is $\mu\tilde g_i$. The transformed couplings are in general complex, which encodes interference effects in the spectral density~\cite{Ben-Asher2026}: a predominantly real coupling produces a Lorentzian-like contribution, a predominantly imaginary coupling an anti-Lorentzian-like contribution, and intermediate cases asymmetric Fano line shapes.

To connect with the figures of merit commonly used for nanophotonic resonators, we also quote the effective mode volume obtained from the usual one-photon energy-density relation, $\tilde V_{\mathrm{eff},i} = \mathrm{Re}\{\tilde\Omega_i\}/(2\varepsilon_0\tilde g_i^2)$. Since $\tilde g_i$ is complex, so is $\tilde V_{\mathrm{eff},i}$, as also found for quasinormal modes of open and lossy resonators~\cite{Wu2021}; we report its modulus in units of $\lambda^3$, where $\lambda = 2\pi c/\mathrm{Re}\{\tilde\Omega_i\}$ is the free-space wavelength. As it is derived from the fit at the emitter position, $\tilde V_{\mathrm{eff},i}$ is a local, fit-derived quantity rather than a directly computed electromagnetic mode volume.



\section{HPhPs in Planar Configurations}
\label{sec:planar}

HPhPs arise from the bulk electromagnetic response of anisotropic polar crystals. We briefly review the properties of HPhPs in hBN that are relevant for the nanocavities considered below, in particular their highly directional, ray-like propagation, which will allow us to identify the cavity resonances from simple geometrical arguments.

\begin{figure}[t]
    \centering
    \includegraphics[width=1.0\linewidth]{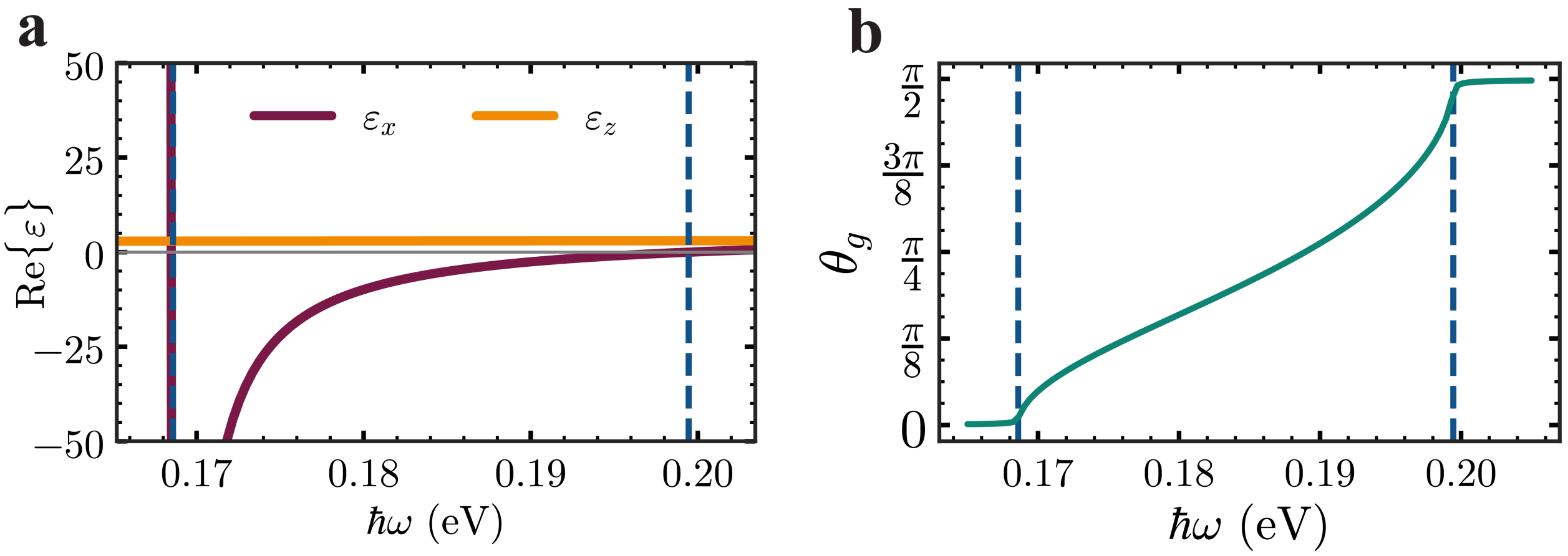}
    \caption{Hyperbolic phonon polaritons in hBN. (a) Real part of the hBN permittivity tensor components within the second Reststrahlen band. (b) Propagation angle $\theta_g$ of the group velocity of high-$k$ HPhPs, measured from the hBN plane, as a function of energy. The dashed lines in both panels indicate $\omega_{TO,x} = 0.1686\,\mathrm{eV}$ and $\omega_{LO,x} = 0.200\,\mathrm{eV}$.}
    \label{fig:fig1}
\end{figure}

In hBN, the different strengths of the in-plane and out-of-plane bonds give rise to distinct optical-phonon responses along the two directions~\cite{Constantinescu2013}. The permittivity tensor is diagonal, $\bar{\bar{\varepsilon}}(\omega) = \text{diag}\left( \varepsilon_{x}(\omega), \, \varepsilon_{x}(\omega), \,\varepsilon_{z}(\omega) \right)$, where $\varepsilon_{x}$ and $\varepsilon_{z}$ are the components in the plane and along the optical axis, respectively. We use the Lorentz-model parameters of Giles et al.~\cite{Giles2018} for isotopically pure $^{11}\mathrm{B}$ hBN, which exhibits substantially lower optical losses than hBN with natural isotope abundance.

Extraordinary waves in the uniaxial medium obey the dispersion relation~\cite{Alvarez-Perez2019}
\begin{equation}
    \label{eq:disp}
    \frac{k_{x}^2}{\varepsilon_{z}} + \frac{k_{z}^2}{\varepsilon_x} = \frac{\omega^2}{c^2}.
\end{equation}
If both permittivity components are positive, this equation describes a conventional elliptical isofrequency contour. If they have opposite signs, the isofrequency contour becomes hyperbolic and supports modes with wave vectors far exceeding the free-space one (Figure~S1a), enabling extreme subwavelength confinement. \autoref{fig:fig1}(a) shows the real parts of the permittivity components of hBN in its second Reststrahlen band (2RB), where the in-plane component is negative while the out-of-plane component is positive.

In the limit $k\gg\omega/c$, the right-hand side of \autoref{eq:disp} can be neglected, and the ratio $k_z/k_x\approx\sqrt{-\varepsilon_x/\varepsilon_z}$ becomes independent of the magnitude of the wave vector. The group velocity $\mathbf{v}_g= \boldsymbol{\nabla}_{\mathbf{k}}\omega$ then points along a fixed direction, at an angle $\theta_g$ from the hBN plane given by $\tan\theta_g(\omega)=\sqrt{-\varepsilon_z(\omega)/\varepsilon_x(\omega)}$. All high-$k$ HPhPs at a given frequency therefore propagate in the same direction, giving rise to the characteristic ray-like propagation of hyperbolic polaritons. As shown in \autoref{fig:fig1}(b), $\theta_g$ increases monotonically with energy across the 2RB, such that the rays become increasingly aligned with the optical axis near its upper edge.

In the same limit, the wave vector and the group velocity become orthogonal (Supplementary Note~S2.1). Equivalently, the high-$k$ field is quasi-electrostatic, since retardation is negligible for $k\gg\omega/c$, but in contrast to isotropic media, where such fields are evanescent, it propagates along the rays. The phase of the field therefore remains essentially constant along a ray, independent of the propagation distance, and the system behaves quasistatically even for dimensions larger than the free-space wavelength. This property is central for the design of hyperbolic cavities: rays reflected at the cavity boundaries return to the emitter without accumulating propagation phase, such that the resonance conditions are determined by the geometry of the ray trajectories alone.

Efficient excitation of these high-$k$ modes requires a source that provides large in-plane momenta. A dipole close to the surface does so naturally: its near field is strongly localized in real space and thus contains a broad distribution of in-plane wave vectors, including evanescent components with $k_\parallel\gg\omega/c$. The closer the dipole is to the surface, the broader this distribution, and the narrower the excited HPhP rays.

The field scattered back by the hBN modifies the spectral density experienced by the dipole. For the emitter-surface separation of $h=1\,\mathrm{nm}$ used throughout this work, the Purcell factor $F_P = J/J_0$, where $J_0$ is the free-space spectral density, reaches values on the order of $10^8$ already above a semi-infinite hBN substrate, and the response of slabs approaches this limit with increasing thickness (Figure~S3b). However, this response is spectrally broad and lacks the discrete, spectrally isolated resonances required for coherent energy exchange with a single molecular vibration. In the finite nanocavities considered below, it provides the background on top of which lateral confinement produces discrete resonances.



\section{Hyperbolic Nanocavities}
\label{sec:cavities}

\begin{figure}[t]
    \centering
    \includegraphics[width=1.0\linewidth]{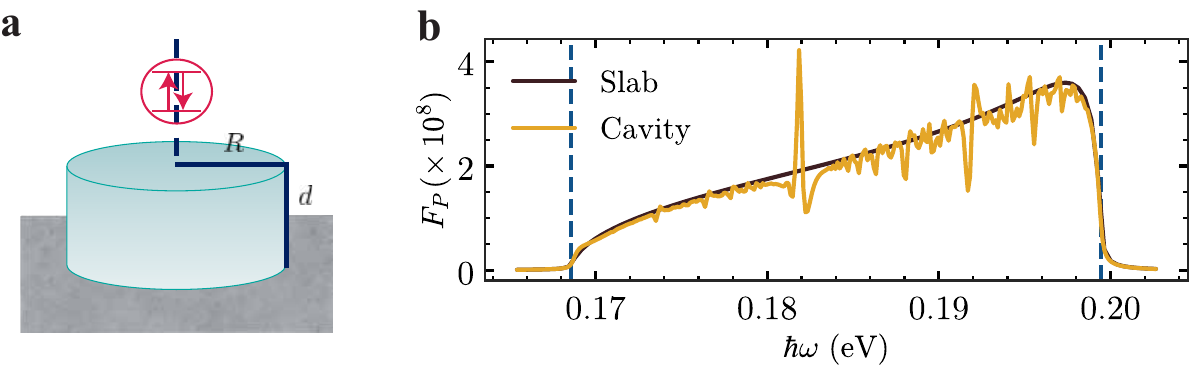}
    \caption{Hyperbolic nanocavities. (a) Schematic of the proposed configuration, consisting of a cylindrical hBN nanocavity with radius $R$ and thickness $d$ on a silica substrate. A molecule with an IR-active vibration is placed on the symmetry axis at a distance $h$ above the cavity surface. (b) Purcell factor $F_P(\omega)=J(\omega)/J_0(\omega)$, where $J_0(\omega)$ is the free-space spectral density, for a $z$-oriented dipole at $h = 1 \,\mathrm{nm}$ above a nanocavity with $d = 25 \,\mathrm{nm}$ and $R = 40 \,\mathrm{nm}$, compared with that above a slab of the same thickness.}
    \label{fig:fig2}
\end{figure}

To create discrete resonances, we laterally confine the HPhPs in a cylindrical hBN nanocavity of thickness $d$ and radius $R$, placed on a silica substrate and surrounded by air ($\varepsilon_b=1$), as shown schematically in \autoref{fig:fig2}(a). Since HPhPs are bulk modes of the anisotropic material, the ray picture introduced above remains applicable inside the cavity, with the additional constraint imposed by the sidewall at $\rho = R$. Instead of propagating away from the source, the rays are now reflected back into the finite cavity volume, which gives rise to localized resonances.

We consider a molecule placed on the symmetry axis at a distance of $h=1\,\mathrm{nm}$ above the cavity surface, with its transition dipole oriented along $z$. In contrast to plasmonic systems~\cite{Esteban2012,Goncalves2020}, nonlocal corrections to the material response remain negligible at this separation (Supplementary Note~S2.4).

\autoref{fig:fig2}(b) shows the Purcell factor for $d = 25 \,\mathrm{nm}$ and $R = 40\,\mathrm{nm}$, together with that of a slab of the same thickness. On top of the slab continuum, the cavity spectrum displays several resonances with asymmetric line shapes within the 2RB. The most pronounced one, at around $0.182\,\mathrm{eV}$, is spectrally narrow and reaches a Purcell factor of about $4 \times 10^8$, significantly larger than typical values in plasmonic systems.

\begin{figure}[t]
    \centering
    \includegraphics[width=1.0\linewidth]{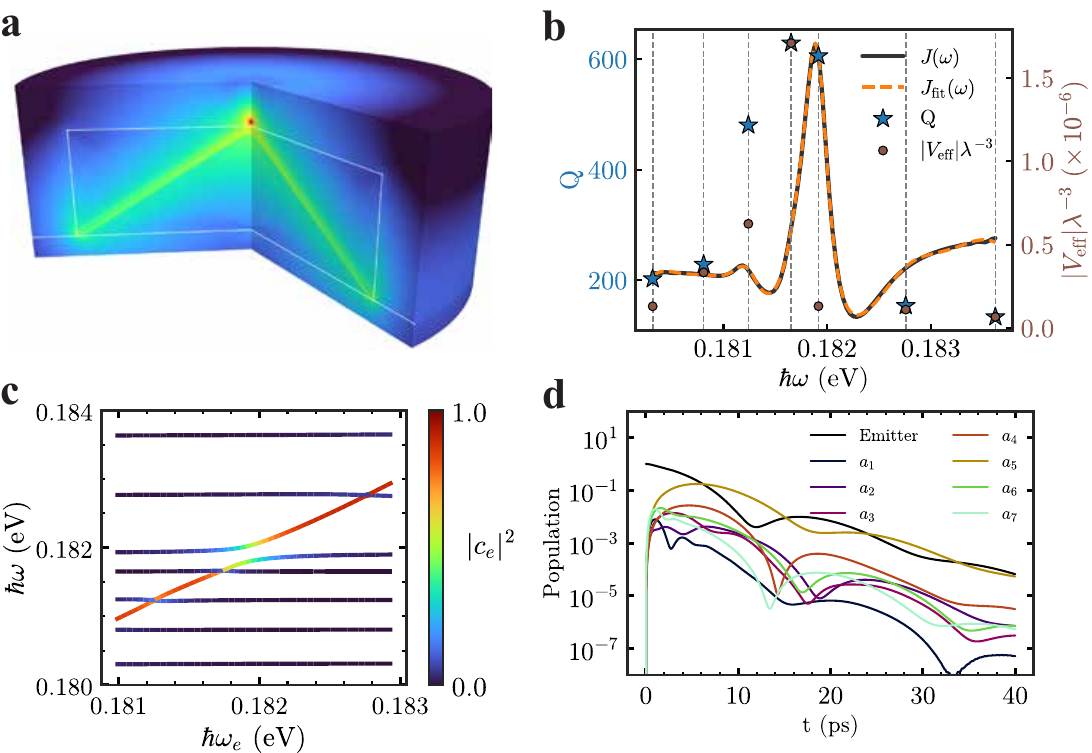}
    \caption{Single-molecule vibrational strong coupling mediated by the conical-hat resonance ($d=25\,\mathrm{nm}$, $R=40\,\mathrm{nm}$). (a) Electric-field intensity profile at $\omega = 0.182\,\mathrm{eV}$, illustrating the ray trajectory associated with the resonance. (b) Few-mode fit of the non-perturbative spectral density in the frequency window surrounding the resonance. (c) Spectrum of the effective non-Hermitian Hamiltonian as a function of the molecular transition frequency for $\mu = 1\,\mathrm{D}$. The color scale indicates the molecular contribution to each eigenstate. (d) Time evolution of the populations of the molecular vibration and the effective modes for $\omega_e = 0.1819\,\mathrm{eV}$, starting from an excited molecular vibration.}
    \label{fig:fig3}
\end{figure}

This resonance corresponds to the simplest closed ray trajectory in the nanocavity. \autoref{fig:fig3}(a) shows the field excited by the dipole at $\omega=0.182\,\mathrm{eV}$: an HPhP ray propagates from the molecule toward the bottom corner of the cavity, is reflected there, and returns to its point of origin, forming a conical-hat-shaped field distribution in three dimensions. Since the reflected field returns to the emitter, it interferes with the field generated directly by the molecule and strongly enhances the spectral density.

The resonance frequency follows from a simple geometrical argument. A ray launched from the center of the top surface hits the bottom corner of the cavity if its propagation angle satisfies $\tan\theta_g(\omega_{\mathrm{res}})=d/R$, with $\tan\theta_g(\omega)=\sqrt{-\varepsilon_z(\omega)/\varepsilon_x(\omega)}$ as derived in \autoref{sec:planar}. For $d=25\,\mathrm{nm}$ and $R=40\,\mathrm{nm}$, this condition yields $\omega_{\mathrm{res}}=0.1819\,\mathrm{eV}$, in excellent agreement with the resonance in \autoref{fig:fig2}(b). Since the condition depends only on the aspect ratio $\mathcal{A}=d/R$, the conical-hat resonance follows lines of constant $\mathcal{A}$ and can be tuned across the 2RB by varying this ratio (Figure~S6a).

The conical-hat resonance is not a single (quasi-)normal mode of the nanocavity. It is more closely related to the pseudomodes found for emitters close to plasmonic nanoparticles~\cite{Tudela2014,Delga2014}, which arise from the collective excitation of many eigenmodes with similar frequencies but a broad range of wave vectors. Owing to the hyperbolic dispersion, this collective excitation does not remain localized near the emitter, but propagates as a narrow ray, whose width is set by the inverse width of the excited wave-vector distribution and thus by the emitter-surface distance. In contrast to plasmonic pseudomodes, which become essentially independent of the particle geometry for emitters close to the surface, the HPhP pseudomodes are shaped by propagation through the cavity volume and are therefore strongly tunable through both the cavity geometry and the emitter position.

We now consider a molecular vibration with transition frequency $\omega_e$ close to the conical-hat resonance and a transition dipole moment of $\mu=1\,\mathrm{D}$, representative of strongly IR-active vibrational modes~\cite{Xiang2020}. The free-space radiative decay of vibrational transitions is negligible due to their low frequency and modest dipole moment, and their relaxation is instead dominated by non-radiative channels such as intramolecular vibrational redistribution (IVR) and energy transfer to the environment. These can be strongly suppressed in small or rigid molecules, through the choice of substrate or matrix, and at cryogenic temperatures~\cite{Mirzaei2026}. Throughout this work, we use an intrinsic non-radiative decay rate of $\gamma_{\text{e}} = 0.04\,\text{meV}$~\cite{Mirzaei2026}, corresponding to a vibrational lifetime of about $16\,\mathrm{ps}$.

Following \autoref{sec:theory}, we fit the spectral density within a frequency window around the resonance with seven effective modes, as shown in \autoref{fig:fig3}(b). The conical-hat resonance is captured mainly by a single effective mode with frequency $\omega_c \equiv \Re\{\tilde\Omega_5\} =0.1819\,\mathrm{eV}$ and decay rate $\kappa_c \equiv -2\Im\{\tilde\Omega_5\} = 0.3\,\mathrm{meV}$, corresponding to a quality factor $Q_c=606$. Its coupling, $\tilde g_c \equiv \tilde g_5 \approx 0.006+0.001i \,\mathrm{V/nm}$ (Figure~S8a), corresponds to an effective mode volume of $|\tilde V_{\mathrm{eff},5}| = 1.2\times10^{-7}\lambda^{3}$, indicating deep subwavelength confinement. The remaining spectral density outside the fitting window (Figure~S7a) produces a Lamb shift of the order of $0.01\,\mathrm{meV}$, such that the molecular transition frequency is only weakly renormalized (Figure~S7b). The residual perturbative decay rate $\gamma_{\mathrm{pert}}$ only reflects the small fit error close to resonance and is negligible (smaller by 2 orders of magnitude) compared to $\gamma_e$. 

The complex coupling reflects the asymmetric, interference-induced line shape of the resonance. Schematically, the spectral density at the emitter can be decomposed as $J(\mathbf r,\omega) \propto \text{Im}\{G_{\text{free},zz} + G_{\text{bg},zz} + G_{\text{refl},zz}\}$, where $G_{\text{bg},zz}$ is the broadband response inherited from the slab and $G_{\text{refl},zz}$ is the contribution of rays reflected by the cavity boundary. The cavity thus modifies the self-interaction of the molecule by feeding back a reflected HPhP field on top of the slab response, and the interference between the two produces the asymmetric line shape.

Whether this interference is constructive or destructive depends on the phase acquired upon reflection, and thus on the boundary at which the HPhPs are reflected. As a benchmark, we considered a geometry analogous to that investigated in Ref.~\citenum{HerzigSheinfux2024}, in which the hBN is not patterned. Instead, an extended hBN slab is placed on a gold film containing an aperture, which we model as a cylindrical air hole, with the molecule again located on the symmetry axis at $h=1\,\mathrm{nm}$ above the hBN surface (Supplementary Note~S5.5). Since the HPhP dispersion of hBN on gold differs strongly from that of hBN on air, the rays launched by the molecule are partially reflected at the edge of the aperture, which thus plays the role of the cavity sidewall. The spectral density again displays a geometry-induced feature at the frequency for which the hyperbolic ray returns from this edge to the emitter, but with an antiresonance line shape dominated by a dip below the background, signaling predominantly destructive interference between the reflected field and the field generated directly by the molecule. In the few-mode description, this feature is reproduced by effective modes whose couplings have imaginary parts exceeding their real parts, reflecting the anti-Lorentzian-like character of the line shape. Consequently, a molecular vibration with a dipole moment of $1\,\mathrm{D}$ does not exhibit a Rabi splitting or coherent energy exchange with the polaritonic modes in this geometry. The interference of the reflected HPhP field at the emitter can therefore be controlled by engineering the interface at which the rays are reflected.

Since the conical-hat resonance arises from the constructive interference of HPhP rays at the molecule position, rather than from a single mode with a well-defined frequency and spatial profile, an expansion of the Green's function in the quasinormal modes (QNMs) of the nanocavity would require a very large number of modes to reproduce the features observed here. The few-mode construction instead provides a compact representation tailored to the spectral density relevant for the emitter dynamics, in which the dominant effective mode represents the pseudomode-like conical-hat resonance rather than an individual QNM\@.

Diagonalizing the effective non-Hermitian Hamiltonian of the coupled system as a function of the molecular transition frequency yields the spectrum shown in \autoref{fig:fig3}(c), where the color scale indicates the molecular contribution to each eigenstate. A clear avoided crossing appears at $\omega_e \approx 0.1819\,\mathrm{eV}$, with a splitting of $\Delta E \approx 0.24\,\mathrm{meV}$. At resonance, the two hybrid eigenstates consist almost exclusively of the molecular vibration and the dominant effective mode, with nearly equal weights (Figures~S8b,c). The system is therefore well described by a Jaynes-Cummings model with a single mode and a complex light-matter coupling~\cite{Ben-Asher2026}.

In this single-mode model, the splitting at zero detuning is $\Delta E = \Re\left[\sqrt{\tilde\Omega_R^2 - (\kappa_c - \gamma_e)^2/4}\right]$, where $\tilde\Omega_R = 2 \mu \tilde g_c$ is the complex vacuum Rabi frequency. With the parameters of the dominant effective mode, this gives $\Delta E = 0.22\,\mathrm{meV}$, in close agreement with the $0.24\,\mathrm{meV}$ obtained from the full few-mode model. This single-mode estimate neglects the broadband contribution of the remaining effective modes, which the close agreement with the full few-mode result shows to be small. The onset of strong coupling can be quantified by the dimensionless criteria (extended to complex coupling and valid when $\Re\tilde{\Omega}_R \gg |\!\Im\tilde{\Omega}_R|$, see Supplementary Note~S4.4)~\cite{Rider2022}
\begin{equation}
    C_1 = \frac{\Re(\tilde\Omega_R)}{|\kappa_c-\gamma_e|/2},
    \qquad
    C_2 = \frac{\Re(\tilde\Omega_R)}{\sqrt{(\kappa_c^2+\gamma_e^2)/2}}.
\end{equation}
At zero detuning, $C_1>1$ marks the appearance of a nonzero splitting of the real parts of the two eigenvalues, while the more stringent condition $C_2>1$ requires this splitting to exceed the linewidths of the hybrid states, such that it can be resolved spectroscopically~\cite{Rider2022}. Here, we find $C_1 \approx 1.96$ and $C_2 \approx 1.19$, so that both criteria are fulfilled.

Finally, we solve the master equation, \autoref{eq:meqtime}, starting from an excited molecular vibration. As shown in \autoref{fig:fig3}(d), the populations display several cycles of coherent energy exchange between the molecule and the dominant effective mode before dissipation damps the oscillations, while the remaining effective modes acquire only small transient populations. Together with the resolvable splitting, these coherent oscillations confirm that the conical-hat resonance supports single-molecule vibrational strong coupling for realistic molecular parameters.

\begin{figure}[t]
    \centering
    \includegraphics[width=0.5\linewidth]{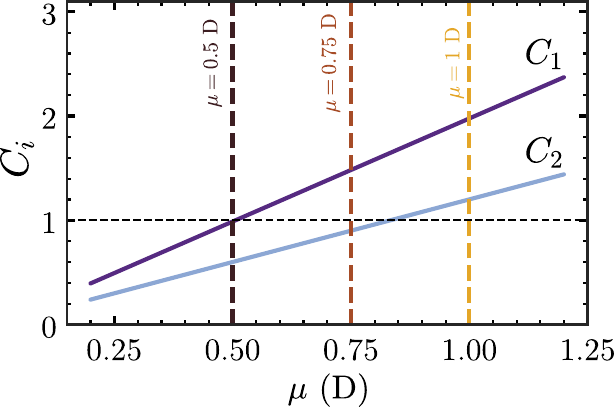}
    \caption{Strong-coupling criteria $C_1$ and $C_2$ as a function of the molecular transition dipole moment $\mu$ for the conical-hat resonance of a nanocavity with $d=25\,\mathrm{nm}$ and $R=40\,\mathrm{nm}$, evaluated within the single-mode model with $\gamma_e = 0.04\,\mathrm{meV}$, $\kappa_c = 0.3\,\mathrm{meV}$, and $\mathrm{Re}\{\tilde g_c\} = 0.006\, \mathrm{V/nm}$. The vertical dashed lines indicate the dipole moments discussed in the text.}
    \label{fig:fig4}
\end{figure}

Strong coupling can also be reached for weaker vibrational transitions. In \autoref{fig:fig4}, we use the criteria $C_1$ and $C_2$ to estimate the minimum dipole moments required for the onset of eigenvalue splitting and for a spectroscopically resolvable splitting, respectively. For the conical-hat resonance, we obtain the threshold values $\bar\mu_1=0.52\,\mathrm{D}$ and $\bar\mu_2=0.83\,\mathrm{D}$. For $\bar\mu_1<\mu<\bar\mu_2$, the eigenvalues still split, but the splitting is not expected to be spectroscopically resolvable. Consistent with this distinction, for $\mu=0.75\,\mathrm{D}$, the population dynamics still exhibit visible coherent oscillations, with a splitting of approximately $0.14\,\mathrm{meV}$, whereas for $\mu=0.5\,\mathrm{D}$, the avoided crossing disappears and coherent oscillations are no longer observed (Figures~S8e,f).



\section{Cavity-Size Control of Strong Coupling}
\label{sec:size}

In the previous section, we focused on the conical-hat resonance because of its simple ray interpretation, well-defined field profile, and pronounced enhancement of the spectral density. It is, however, only one of many resonances supported by the nanocavities.

\begin{figure}[t]
    \centering
    \includegraphics[width=1.0\linewidth]{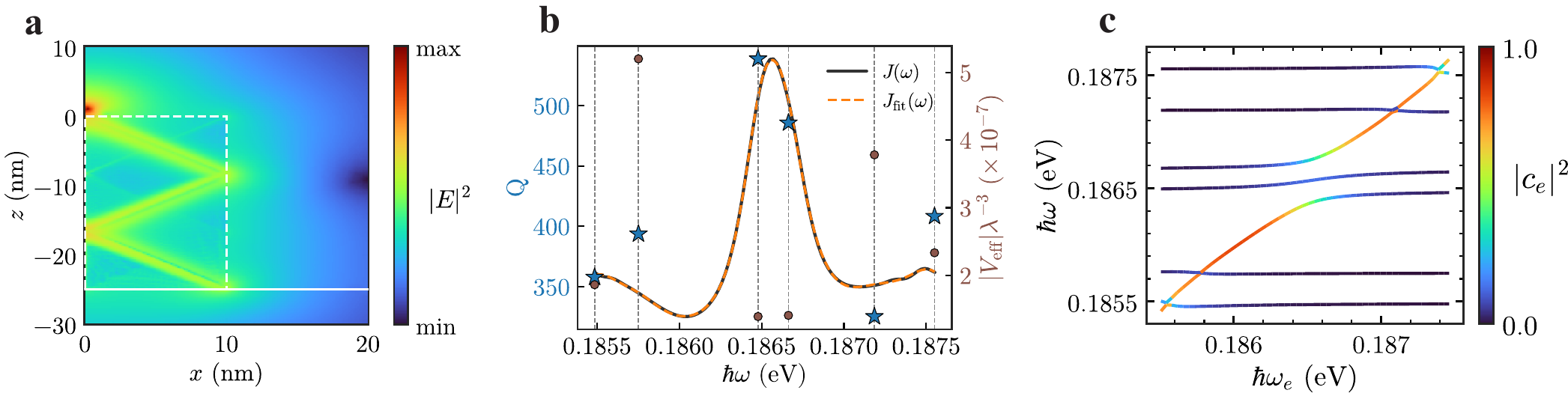}
    \caption{Strong coupling to a hyperbolic resonance in a nanocavity with $d = 25\,\mathrm{nm}$ and $R = 10\,\mathrm{nm}$. (a) Electric-field intensity profile in the $xz$ plane at $\omega = 0.1865\,\mathrm{eV}$. (b) Few-mode fit of the non-perturbative spectral density. (c) Spectrum of the effective non-Hermitian Hamiltonian as a function of the molecular transition frequency for $\mu = 1\,\mathrm{D}$. The color scale indicates the molecular contribution to each eigenstate.}
    \label{fig:fig5}
\end{figure}

For fixed thickness $d$ and varying radius $R$, the spectral density displays a large number of interference-induced resonances, corresponding to different HPhP trajectories and field distributions (Figure~S6b). Many of them can be anticipated from geometrical considerations, without explicitly calculating the field, by identifying ray trajectories that close on the emitter position. Since the HPhP rays accumulate essentially no propagation phase, the cavity dimensions provide two largely independent knobs: the aspect ratio selects the ray trajectory and thereby the resonance frequency, while the absolute cavity size controls the field confinement at the emitter and thereby the light-matter coupling strength.

As an example, we consider a nanocavity with the same thickness as before, $d=25\,\mathrm{nm}$, but a smaller radius, $R=10\,\mathrm{nm}$. Its spectral density displays several resonances with distinct field profiles, including the conical-hat resonance (Figure~S9a). We focus on a resonance at $\omega=0.1866\,\mathrm{eV}$, whose field profile is shown in \autoref{fig:fig5}(a). Here, the ray launched by the emitter first hits the cavity sidewall, is reflected toward the opposite bottom corner, and returns to the emitter along the same path. The corresponding geometrical condition, $\tan\theta_g(\omega_{\mathrm{res}})=d/(3R)$, gives $\omega_{\mathrm{res}}=0.1863\,\mathrm{eV}$, in close agreement with the numerically observed resonance.

At this frequency, the spectral density reaches $\hbar J=0.065\,\mathrm{meV}$, compared with $\hbar J=0.045\,\mathrm{meV}$ at the conical-hat resonance of the $R=40\,\mathrm{nm}$ cavity. Like the conical-hat resonance, this resonance is formed by many underlying cavity modes, and here the few-mode fit requires two effective modes to reproduce it accurately (\autoref{fig:fig5}(b)), with quality factors $Q=480$ and $540$ and nearly identical complex couplings $\tilde g\approx0.006-0.0004i\,\mathrm{V/nm}$. The molecule couples to the bright superposition of these two modes, with a collective coupling of approximately $\sqrt{2}\,\mu\tilde g$, corresponding to an effective mode volume of the resonance of $|\tilde V_{\mathrm{eff}}|\approx0.7\times10^{-7}\lambda^{3}$, smaller than that of the conical-hat resonance. Neglecting losses, this gives an expected splitting of about $0.37\,\mathrm{meV}$, close to the $0.39\,\mathrm{meV}$ obtained from the full effective Hamiltonian in \autoref{fig:fig5}(c). The corresponding dynamics display coherent oscillations between the molecular vibration and the effective modes (Figure~S9b).

One may expect that further reducing the cavity size enhances the light-matter interaction even more. To explore this regime, we consider an ultrasmall cavity with $d=6.25\,\mathrm{nm}$ and $R=5\,\mathrm{nm}$, corresponding to about $19$ atomic layers of hBN. While its fabrication would be challenging, it serves as a proof-of-principle example of how reducing the cavity dimensions decreases the effective mode volume and thereby enhances the light-matter interaction.

This ultrasmall cavity differs qualitatively from the ray picture developed above. No prominent resonance appears at the frequency predicted by the conical-hat condition. Instead, the spectral density is dominated by nearby resonances with different field distributions, which need not correspond to trajectories that return exactly to the emitter (Figure~S10a). Here, the width of the HPhP rays, which is set by the emitter-surface distance, becomes comparable to the cavity dimensions, such that the ray picture no longer accurately describes the resonant fields. Nevertheless, the fields of these resonances interfere constructively at the emitter position, producing Purcell factors on the order of $10^9$.

The dominant resonance is accurately reproduced by a single effective mode with $Q\approx650$ and $|\tilde V_{\mathrm{eff}}|\approx3\times10^{-8}\lambda^{3}$, showing that resonances with simultaneously high quality factors and small effective mode volumes persist beyond the regime in which the ray picture applies. For $\mu=1\,\mathrm{D}$, the spectrum displays a splitting of $0.55\,\mathrm{meV}$ at $\omega_e=0.1895\,\mathrm{eV}$, and the dynamics show pronounced Rabi oscillations (Supplementary Note~S5.4).

The reduced effective mode volumes of the smaller cavities, and the correspondingly larger coupling strengths, allow strong coupling to be reached for molecular transitions with smaller dipole moments. Applying the criterion $C_1$ introduced above, we find that the onset of strong coupling in these cavities extends down to transition dipole moments of approximately $0.2\,\mathrm{D}$. Tuning the cavity dimensions thus provides a route to controlling the interaction strength and extending vibrational strong coupling to molecules with comparatively weak infrared transitions.



\section{Discussion}
\label{sec:discussion}

The resonance frequencies are set by the cavity geometry and are therefore sensitive to the cavity dimensions: for the conical-hat resonance, changing the radius by about $1\,\mathrm{nm}$ shifts the resonance by roughly one linewidth, and a comparable shift results from a change of the thickness by about $1\%$, i.e., $\Delta d\approx0.25\,\mathrm{nm}$ for $d=25\,\mathrm{nm}$. Variations that scale $R$ and $d$ by the same factor, in contrast, leave the aspect ratio and hence the resonance frequency essentially unchanged. This sensitivity does not require sub-nanometer fabrication control: such small variations mainly shift the resonance frequency, and suitable nanocavities can be selected from a fabricated ensemble. The cavity thickness is fixed by the number of hBN layers and can be determined precisely, and the geometrical resonance condition provides a simple design rule for this selection. Since the relevant resonances are formed by high-$k$ HPhPs and are essentially dark in the far field, their spectroscopic characterization requires near-field techniques such as scattering-type near-field optical microscopy (SNOM)~\cite{HerzigSheinfux2024}, in which a tip placed above the cavity center probes a response closely related to the spectral density experienced by the molecule. A more demanding requirement is the position of the molecule, which must lie close to the cavity symmetry axis to couple efficiently to the geometrically selected resonances considered here; the scanning-probe techniques needed for this purpose would also enable the near-field characterization of the cavities. The achievable quality factors are ultimately limited by phonon damping in hBN, making isotopically enriched material highly advantageous~\cite{Giles2018,Guo2025}, and operation at low temperature could further reduce the damping~\cite{Ni2021}. On the molecular side, the strong-coupling regime identified here remains accessible for vibrational lifetimes as short as a few picoseconds.

From a methodological perspective, the combination of macroscopic QED with the few-mode framework is essential for this analysis. The spectral densities are dominated by asymmetric, interference-induced resonances on top of a broad hyperbolic background, which cannot be captured by a set of independent Lorentzian modes. The few-mode approach reproduces these structured spectral densities with a small number of effective modes, provides direct access to the figures of merit governing the light-matter interaction, and yields a compact physical picture of the resonances responsible for strong coupling.



\section{Conclusions}
\label{sec:conclusions}

In summary, we have proposed hyperbolic nanocavities as a platform for single-molecule vibrational strong coupling. Extended hBN slabs already strongly enhance the spectral density experienced by a nearby molecule because of the extreme confinement of HPhPs, but their response is spectrally broad. Laterally confining the HPhPs in finite nanocavities produces discrete resonances that strongly enhance the light-matter interaction. Their frequencies follow from simple ray trajectories of the HPhPs and can be tuned through the cavity aspect ratio, while the absolute cavity size controls the light-matter coupling strength.

For the conical-hat resonance of a cavity with $d=25\,\mathrm{nm}$ and $R=40\,\mathrm{nm}$, we find $Q\approx600$ and $|\tilde V_{\mathrm{eff}}|\approx10^{-7}\lambda^{3}$, leading to a spectroscopically resolvable splitting of $0.24\,\mathrm{meV}$ for a transition dipole moment of $1\,\mathrm{D}$, while smaller cavities extend the onset of strong coupling to dipole moments of approximately $0.2\,\mathrm{D}$. This regime is compatible with experimentally accessible material parameters, although its realization will require accurate molecular positioning and low-loss isotopically enriched hBN.

Single-molecule vibrational strong coupling would make it possible to study how strong coupling modifies individual molecules, and may ultimately open routes toward the local control of vibrational dynamics and molecular energy flow. It would furthermore enable quantum optics with molecular vibrations in the mid-infrared. More broadly, our results show that resonances in hyperbolic cavities can be engineered to realize strong light-matter interactions in deeply subwavelength volumes, establishing them as a promising platform for vibrational quantum nanophotonics.




\section{Data Availability}
\label{sec:data_avail}

The numerical data and Jupyter notebooks supporting the findings of this study are publicly available at \url{https://github.com/stvrath/HyperNano4VSC}.



\section{Acknowledgements}

S.A. and O.J.F. acknowledge funding from the Swiss National Science Foundation (project CRSll5\_216629). F.J.G.-V.\ and J.F.\ acknowledge financial support by the Spanish Ministerio de Ciencia y Universidades-Agencia Estatal de Investigación through Grants No.\ PID2024161142NB-I00, and No.\ CEX2023-001316-M through the María de Maeztu program for Units of Excellence in R\&D\@, as well as by Programa Fundamentos FBBVA through the Grant No.\ EIC24-1-17304.
S.A. would also like to acknowledge insightful discussions with Carlos J. Sánchez Martínez and Maksim Lednev. 


\begin{suppinfo}

    We provide supplementary information on the semi-analytical expression for the spectral density in planar geometries, additional discussion of hyperbolic phonon polaritons, details of the electromagnetic calculations, a description of the few-mode framework and its implementation and supplementary results supporting the main text.

\end{suppinfo}

    \bibliography{bibliography}






\end{document}